\documentstyle[12pt]{article}
\catcode`\@=11
\catcode`\@=11
\def\eqnarray{\let\@currentlabel=\theequation\refstepcounter{equation}
    \global\@eqnswtrue
    \global\@eqcnt\z@\tabskip\@centering\let\\=\@eqncr
    $$\halign to \displaywidth\bgroup\@eqnsel\hskip\@centering
      $\displaystyle\tabskip\z@{##}$&\global\@eqcnt\@ne
       \hfil${{}##{}}$\hfil
      &\global\@eqcnt\tw@ $\displaystyle\tabskip\z@{##}$\hfil
       \tabskip\@centering&\llap{##}\tabskip\z@\cr}
\def\lefteqn#1{\hbox to 4\arraycolsep{$\displaystyle #1$\hss}}
\long\def\@makefntext#1{\parindent 0cm\noindent
\hbox to 1em{\hss$^{\@thefnmark}$}#1}
\newcommand{\beq}{\begin{equation}}
\newcommand{\eeq}{\end{equation}}
\newcommand{\bea}{\begin{eqnarray}}
\newcommand{\eea}{\end{eqnarray}}
\begin{document}
%
%
%
%
\def\citen#1{%
\edef\@tempa{\@ignspaftercomma,#1, \@end, }
\edef\@tempa{\expandafter\@ignendcommas\@tempa\@end}%
\if@filesw \immediate \write \@auxout {\string \citation {\@tempa}}\fi
\@tempcntb\m@ne \let\@h@ld\relax \let\@citea\@empty
\@for \@citeb:=\@tempa\do {\@cmpresscites}%
\@h@ld}
%
\def\@ignspaftercomma#1, {\ifx\@end#1\@empty\else
   #1,\expandafter\@ignspaftercomma\fi}
\def\@ignendcommas,#1,\@end{#1}
%
%
\def\@cmpresscites{%
 \expandafter\let \expandafter\@B@citeB \csname b@\@citeb \endcsname
 \ifx\@B@citeB\relax 
    \@h@ld\@citea\@tempcntb\m@ne{\bf ?}%
    \@warning {Citation `\@citeb ' on page \thepage \space undefined}%
 \else
    \@tempcnta\@tempcntb \advance\@tempcnta\@ne
    \setbox\z@\hbox\bgroup 
    \ifnum\z@<0\@B@citeB \relax
       \egroup \@tempcntb\@B@citeB \relax
       \else \egroup \@tempcntb\m@ne \fi
    \ifnum\@tempcnta=\@tempcntb 
       \ifx\@h@ld\relax 
          \edef \@h@ld{\@citea\@B@citeB}%
       \else 
          \edef\@h@ld{\hbox{--}\penalty\@highpenalty \@B@citeB}%
       \fi
    \else   
       \@h@ld \@citea \@B@citeB \let\@h@ld\relax
 \fi\fi%
 \let\@citea\@citepunct
}
%
\def\@citepunct{,\penalty\@highpenalty\hskip.13em plus.1em minus.1em}%
%
%
\def\@citex[#1]#2{\@cite{\citen{#2}}{#1}}%
%
%
\def\@cite#1#2{\leavevmode\unskip
  \ifnum\lastpenalty=\z@ \penalty\@highpenalty \fi 
  \ [{\multiply\@highpenalty 3 #1
      \if@tempswa,\penalty\@highpenalty\ #2\fi 
    }]\spacefactor\@m}
\let\nocitecount\relax  
%
\def\IR{{\hbox{{\rm I}\kern-.2em\hbox{\rm R}}}}
\def\IH{{\hbox{{\rm I}\kern-.2em\hbox{\rm H}}}}
\def\IC{{\ \hbox{{\rm I}\kern-.6em\hbox{\bf C}}}}
\def\II{{\ \hbox{{\rm I}\kern-.6em\hbox{\bf I}}}}
\def\IZ{{\hbox{{\rm Z}\kern-.4em\hbox{\rm Z}}}}
\def \um {{1\over 2}}
\def \ga {\gamma}
\def \om {\omega}
\def \th {\theta}
\def \ep {\epsilon}
\def \la {\lambda}
\def \a {\alpha}
\def \La {\Lambda}
\def \Da {\Delta}
\def\rref#1{(\ref{#1})}
\def\be{\begin{equation}}
\def\ee{\end{equation}}
\def \Ga {\Gamma}
\def \t {\tau}
\def \sig {\sigma}
\def \b {\beta}
\def \g {\gamma}
\def\II{{\hbox{{\rm I}\kern-.2em\hbox{\rm I}}}}
\begin{titlepage}
\vspace{.5in}
\begin{flushright}
DFTT/59/99\\
IST/DM/30/99\\
November 1999\\
gr-qc/9911005\\
\end{flushright}
\begin{center}
{\Large\bf
Quantum Holonomies\\ [1ex]
in (2+1)-Dimensional Gravity}\\
\vspace{.4in}
{J.\ E.~N{\sc elson}\footnote{\it email: nelson@to.infn.it}\\
       {\small\it Dipartimento di Fisica Teorica }\\
       {\small\it Universit\`a degli Studi di Torino}\\
       {\small\it via Pietro Giuria 1, 10125 Torino}\\{\small\it
Italy}}\\
\vspace{1ex}
{\small and}\\
\vspace{1ex}
{R.\ F.~P{\sc icken}\footnote{\it  email: picken@math.ist.utl.pt}\\
{\small\it Departamento de Matem\'{a}tica and Centro de
Matem\'{a}tica Aplicada}\\
{\small\it  Instituto Superior T\'{e}cnico}\\
{\small\it Avenida Rovisco Pais, 1049-001 Lisboa}\\{\small\it
Portugal}}\\
\end{center}
\vspace{.5in}
\begin{center}
\begin{minipage}{4.8in}
\begin{center}
{\large\bf Abstract}
\end{center}
{\small
We describe an approach to the quantisation of
(2+1)--dimensional gravity
with topology $\IR \times T^2$ and negative cosmological constant,
which uses two quantum holonomy matrices satisfying a
$q$--commutation relation. Solutions of diagonal and
upper--triangular form are constructed, which in the latter case exhibit
additional, non--trivial {\it internal} relations for each
holonomy matrix. Representations are constructed and a group of
transformations - a quasi--modular group - which preserves this
structure, is presented.}
\begin{flushleft}
P.A.C.S.04.60Kz, 02.10.Tq, 02.20.-a\\
\end{flushleft}
\end{minipage}
\end{center}
\end{titlepage}
\addtocounter{footnote}{-2}
\section{Introduction \label{sec1}}
It is known \cite{NRZ,NR1} that the phase space of
(2+1)-dimensional gravity with topology $\IR\!\times\!T^2$ and
negative cosmological constant $\Lambda$ can be
described in terms of six gauge-invariant (normalised) traces of
 $\hbox{SL}(2,\IR)$ holonomies
$~ T_1^{\pm},  T_2^{\pm}$, and  $T_{12}^{\pm}$,
which satisfy the non--linear Poisson bracket algebra
\beq
\{T_1^{\pm},T_2^{\pm}\}=\mp{{\sqrt {-\Lambda}}\over 4}(T_{12}^{\pm}-
 T_1^{\pm}T_2^{\pm}),
\label{b7}
\eeq
and~cyclical~permutations~of $T_i^{\pm}, i=1,2,12.$ The six traces
$T_i^\pm$ that satisfy \rref{b7} provide an overcomplete
description of the spacetime geometry of $\IR\!\times\!T^2$, since
the cubic polynomials
\beq
F^{\pm}=1-(T_1^{\pm})^2-(T_2^{\pm})^2-(T_{12}^{\pm})^2 +
 2 T_1^{\pm}T_2^{\pm}T_{12}^{\pm} .  \label{b9} \eeq have vanishing
Poisson brackets with all of the traces $T_i^{\pm}$ and may be set
to zero. This corresponds to imposing the classical
$\hbox{SL}(2,\IR)$ Mandelstam identities.  The traces $T_i^\pm$ can
be represented classically as \cite{cn1}
\beq
T_1^\pm = \cosh{r_1^\pm\over2} , \quad T_2^\pm =
\cosh{r_2^\pm\over2} , \quad
T_{12}^\pm = \cosh{(r_1^\pm+r_2^\pm)\over2} , \label{cc6}
\eeq
and satisfy the constraints
$F^{\pm}=0$. They will satisfy the algebra \rref{b7} provided the
global parameters $r^{\pm}$ satisfy
\beq
\{r_1^\pm,r_2^\pm\}=\mp \sqrt{-\Lambda}, \qquad \{r^+_a,r^-_b\}=0 .
\label{a1}
\eeq
Quantisation of \rref{a1} would then give
\beq
[\hat r_1^\pm, \hat r_2^\pm] = \mp {i\hbar \sqrt{-\Lambda}} .
\label{dc1}
\eeq
Previous quantisations \cite{cn1,cn2} have
concentrated entirely on the traces $T_i^{\pm}$ and their
representation \rref{cc6}. In this letter we observe that we may
regard the quantised traces $\hat T_i^\pm$ as traces of diagonal
operator-valued holonomy matrices $\hat T_i^{\pm}= \um {\rm tr}
\hat U_i^{\pm}$, $i=1,2$, $\hat T_{12}^{\pm}= \um {\rm tr}(\hat
U_1^{\pm} \hat U_2^{\pm})$, where (for the (+)  matrices, we drop
the superscript) the matrices $\hat U_i$ have the form \be \hat U_i
= \left(\begin{array}{clcr}e^{{\hat r}_i \over 2}&0\\0& e^{-{{\hat
r}_i \over 2}} \end{array}\right)\quad = \exp{({{\hat r_i \sig_3}
\over 2})} \label{diag} \ee where $\sig_3$ is one of the Pauli
matrices. Now, from \rref{dc1} and the identity
$$ e^{\hat X}
e^{\hat Y}= e^{\hat Y} e^{\hat X} e^{[ \hat X, \hat Y ]},
$$
when
$[ \hat X, \hat Y ]$ is a $c$--number one finds that the
matrices \rref{diag} satisfy, {\it by both matrix and operator
multiplication}, the $q$--commutation relation:
\beq
\hat U_1 \hat U_2 = q \hat U_2 \hat U_1,
\label{fund}
\eeq
with
\beq
q=\exp ({{- i \hbar \sqrt{-\Lambda}} \over 4})
\label{q}
\eeq
i.e. a deformation of the classical equation stating that the holonomies
commute.

Equations of the form \rref{fund} appear abundantly in the
quantum group
and quantum geometry literature, e.g. as the defining relation for
the quantum plane \cite{man}, or the non-commutative 2-dimensional torus
\cite{rieff}, but normally
the symbols $\hat U_1$ and $\hat U_2$ stand for scalar operators, as
opposed to $2\times 2$ matrices with operator
entries. The only case similar to ours that we are aware of is Majid's
construction of {\em braided
matrices}  \cite[Section 10.3]{maj}. However, these matrices differ
substantially in the structure
of the internal relations amongst matrix entries, which we
discuss shortly.

In this letter we base our approach on the fundamental equation
\rref{fund}. Instead of representing the algebra of traces
\rref{b7}, we find representations of matrices $\hat U_1$ and $\hat
U_2$ satisfying equation \rref{fund} that generalise the choices
\rref{diag}, for a general $q$--parameter. This constitutes a new
approach to quantisation that is consistent with previous
approaches for this model \cite{NR1,cn1,cn2}, namely a deformation
of classical holonomies that consequently satisfy a
$q$--commutation relation. In this approach the gauge-invariance of
the traces is replaced by the gauge-covariance of \rref{fund} under
the replacements $\hat U_i\rightarrow g^{-1}\hat U_ig$, $i=1,2$ for
$g\in \hbox{SL}(2,\IR )$ an ordinary, i.e. not operator-valued,
matrix. We argue that working directly with the matrices $U_i$,
rather than with the indirect information contained in their
traces, gives a clearer insight into the structure of the phase
space, both classically and after quantisation.

A more detailed account of the results presented here is given in
\cite{jr21}. The matrices $\hat U_1$ and $\hat U_2$ determine a new
quantum--group--like structure, which is studied from the algebraic
perspective in \cite{jrmath}. The description of the classical
phase space in terms of pairs of matrices $U_i$ is given in
\cite{jrmod}.

Our material is organised as follows. In Section \ref{sec2} we
study the algebraic properties of upper--triangular operator
matrices which satisfy \rref{fund}. In Section \ref{sec3} we find
representations which require both non--trivial internal relations
and mutual $q$--relations. In Section \ref{sec4} we introduce a set
of quasi--modular transformations which generate new quantum
matrices and which leave the fundamental relation \rref{fund}
invariant. Our results are summarised in Section \ref{sec5}.

\section{Upper-Triangular Matrices\label{sec2}}

The classical counterpart to equation \rref{fund} is the statement
that the two matrices $U_1$ and $U_2$ commute. The classical phase
space consists of pairs of commuting $\hbox{SL}(2,\IR)$ matrices,
identified up to simultaneous conjugation by elements of
$\hbox{SL}(2,\IR)$.  This classical phase space is studied in full
detail elsewhere \cite{jrmod}.  It consists of not only sectors
where both matrices are diagonalisable, but also sectors where both
are non-diagonalisable but can be simultaneously conjugated into
upper triangular form, as well as other sectors besides these. Here
we analyse pairs of upper triangular quantum matrices, as a first
step towards understanding the quantum version of the classical
phase space. The algebraic analysis of the classical case, in terms
of the eigenvalues and eigenspaces of the two matrices, does not
carry over in any straightforward way to quantum matrices. For
instance, an upper--triangular matrix with two distinct diagonal
entries can be diagonalised as an ordinary matrix, but this is not
in general true if the matrix has non--commuting entries, as will
be the case here. The traced holonomy variables do not distinguish
between diagonal and upper--triangular quantum matrices, another
motivation for using the matrices themselves as variables.

We first study \rref{fund} from an algebraic point of view. These
aspects are interesting in their own right and are studied in
\cite{jrmath}. Here we limit ourselves to a brief discussion.
Consider two upper--triangular matrices of the form
\be
V_i=\left(\begin{array}{clcr}{\a_i}&{\b_i}\\0& {\g_i}
\end{array}\right), \qquad i=1,2 \label{matrv}
\ee
whose entries
$\alpha_i,\beta_i,\gamma_i$ are elements of a non-commutative
algebra, with $\alpha_i, \gamma_i$ invertible. We require that the
matrices \rref{matrv} satisfy the fundamental matrix relation
\be
V_1V_2=qV_2V_1 \label{fundv}
\ee
for some scalar parameter $q$.
Equation \rref{diag} may be regarded as a special case of
\rref{matrv} under the identifications:  $\a_i=e^{{\hat r}_i \over
2}$, $\b_i=0$, and $\g_i=e^{-{{\hat r}_i \over 2}}$ .  The entries
of \rref{matrv} must satisfy, from \rref{fundv}, the mutual
relations
\be
\a_1\a_2=q\a_2\a_1,\quad \g_1\g_2=q\g_2\g_1,
\label{ab1}
\ee
{\rm and}
\be
 \a_1\b_2 +\b_1\g_2=q(\a_2\b_1 +\b_2\g_1).
\label{ab11}
\ee
The relations \rref{ab1} are standard $q$--group
relations where by this phrase we mean, for example, relations of the
form
\bea
 ab=qba, \quad &ac=qca,\quad &ad-da=(q-q^{-1})bc,\nonumber \\
      bc=cb, \quad &bd=qdb, \quad &cd=qdc,
\label{qg}
\eea
satisfied by the non--commuting entries of a $2\times 2$ matrix
\be
U=\left( \begin{array}{cc}a &b\\ c&d\end{array} \right).
\nonumber
\ee
which clearly then commute in the classical limit $q \to 1$. The
relations \rref{ab1} also imply that, for example
\be
\a_1{\a_2}^{-1}=q^{-1}{\a_2}^{-1}\a_1,
\quad {\a_1}^{-1}{\a_2}^{-1}=q{\a_2}^{-1}{\a_1}^{-1},
\label{ab12}
\ee
and similarly for $\g_1,\g_2$. Equation \rref{ab11} is
not of the type \rref{qg}. To solve this equation
we make a simplifying ansatz
(others are possible and are studied in \cite{jr21}):
\be
 \a_1\b_2 =q \b_2\g_1, \quad
 \b_1\g_2=q \a_2\b_1,
\label{pair1}
\ee
and also require the off-diagonal terms in the product $V_1V_2$
(and therefore in $V_2V_1$) to be proportional:
\beq
 \a_1\b_2 =\la \b_1\g_2, \quad \b_2\g_1 =\la \a_2\b_1
\label{hom1}
\eeq
for some non--zero scalar parameter $\la$. It follows from
\rref{ab1}--\rref{hom1} that there are non-trivial
{\it internal} relations for the entries of each matrix, namely:
\be
 \a_i\b_i=\b_i\g_i, \quad i=1,2.  \label{int}
\ee
The internal relations \rref{int}, which also appear in the
operator representations of Section \ref{sec3} are a new feature,
since, in the Poisson algebra of (2+1)--dimensional gravity
\cite{NRZ}, only matrix elements from different holonomies have
non--zero brackets and would therefore not commute on quantisation.
Elements of a single holonomy commute.  We note that the $q$
parameter does not appear in the internal relations \rref{int}
which therefore persist in the classical limit $q\rightarrow 1$,
when the matrices \rref{matrv} commute.

The internal relations \rref{int} are also not standard $q$--group
internal relations \rref{qg}, but are,
however, preserved under matrix multiplication, as for quantum groups.
For example, the product $V_1V_2$ is given by
\be
V_1V_2 = \left(\begin{array}{clcr}{\a_1\a_2}&{(1+\la)
\b_1\g_2}\\0&
{\g_1\g_2}
\end{array}\right)
\label{matrvv}
\ee
whose internal relations are like \rref{int}, by using
\rref{ab1}--\rref{ab11} and \rref{ab12}--\rref{int}. This
feature
is discussed in greater detail in \cite{jrmath}.

In Section \ref{sec3} we study the representations of the matrices
\rref{matrv} in the special case $\a_i ={\ga_i}^{-1}, i=1,2$, with
$\la=1$, though other
choices are possible \cite{jr21,jrmath}. The internal relations
\rref{int} in this case become
\be
\a_i\b_i=\b_i\a_i^{-1},\quad {\a_i}^{-1}\b_i=\b_i\a_i,
\label{ii}
\ee
and the relations \rref{ab1} are equivalent.

The diagonal case, i.e. when $\b_1,\b_2$ are null, is a special case and
satisfies \rref{fundv} with just the relations \rref{ab1}.

\section{Operator Representations \label{sec3}}
We apply the analysis of Section \ref{sec2} to upper--triangular operator
matrices
\be
\hat U_1 = \left(\begin{array}{clcr}{\hat P}_1&{\hat c}\\0&
{{{\hat P}_i}^{-1}}
\end{array}\right) , \quad
\hat U_2 = \left(\begin{array}{clcr}{\hat P}_2&{\hat d}\\0&
{{{\hat P}_2}^{-1}}
\end{array}\right)
\label{matr1}
\ee
(where we have changed notation to emphasise that we are now working with
operators) and require that they satisfy
\rref{fund}
\be
\hat U_1 \hat U_2 = q \hat U_2 \hat U_1.
\label{fund1}
\ee
It follows that the entries of \rref{matr1} must satisfy
\be
{\hat P}_1 {\hat P}_2 = q {\hat P}_2 {\hat P}_1 ,
\quad
{\hat P}_1 {\hat d} +{\hat c} {\hat P}_2^{-1} = q ({\hat P}_2 {\hat c} +
{\hat d} {\hat P}_1^{-1}).
\label {fundp}
\ee
As in Section \ref{sec2} the first of \rref{fundp} is a standard
$q$--group
relation but the second is not. To solve for the second of \rref{fundp}
we first construct a representation of the operators ${\hat P}_i$ that
satisfy the first relation. This allows us to determine the
operators $\hat c,\hat d$.
We represent the operators ${\hat P}_i, i=1,2$  appearing in
\rref{matr1} as acting on functions $\psi(b)$ of a configuration
space variable $b$ by shift and multiplication
\be
{\hat P}_1 \psi(b)=\exp ({d}/{db})\psi(b)=\psi (b+1), \quad
{\hat P}_2 \psi (b)=\exp{(i \hbar b)}\psi (b),
\label{actp12}
\ee
where $\hbar$ is Planck's constant. Equation \rref{actp12} implies
that the $q$--parameter in \rref{fund1}, \rref{fundp} is $q= e^{i\hbar}$,
but clearly $q$ can be given any value by adjusting the definition
\rref{actp12}.

Now, from Section \ref{sec2} the {\it internal} elements
of each of \rref{matr1} should satisfy the unconventional
internal relations \rref{ii} which here read
\be
\hat c {\hat P}_1={{\hat P}_1}^{-1} \hat c, \quad \hat d {\hat P}_2=
{{\hat P}_2}^{-1} \hat d,
\label{cd2}
\ee
and are relations of the type
\be
B e^A  = e^{-A} B
\label{anti}
\ee
which is satisfied when the operators $A$ and $B$ {\it anticommute} i.e.
$A  B  +  B  A  =  0$.  This  suggests that the operators
$\hat c, \hat d$ should
anticommute with the operator $ \hat b$, given by $\psi(b) \mapsto b~
\psi(b)$, and therefore also with ${d}/{db}$, which occur in
\rref{actp12}.
Indeed, one solution of \rref{cd2} is given by setting $\hat c$ and
$\hat d$  equal to the parity operator $T$, which acts on $\psi(b)$ as
\be
T\psi(b) = \psi(-b),
\label{t}
\ee
anticommutes with the operators $\hat b$ and ${d}/{db}$
\be
T \hat b T= - \hat b, \quad T {d}/{db}T= - {d}/{db},
\label{tid}
\ee
and satisfies
\be
T{\hat P}_i =  {{\hat P}_i}^{-1}  T, \quad {\hat P}_i T= T {{\hat P}_i}^{-1}
, \quad  i=1,2
\label{ppt}
\ee
which follows from \rref{tid} and \rref{actp12} but can also be
checked directly, e.g.
\bea
{\hat P}_1 T\psi(b)&=&{\hat P}_1 \psi(-b)= \psi(-b-1)=T \psi(b-1)=
T {{\hat P}_1}^{-1} \psi(b)\nonumber\\
{\hat P}_2 T \psi(b)&=&{\hat P}_2 \psi(-b)=e^{i \hbar b}\psi(-b)=
T e^{-i \hbar b}\psi(b)=T {{\hat P}_2}^{-1}\psi(b).
\label{pt}
\eea
The simplest solution for the off--diagonal operators $\hat c,
\hat d$ that satisfies \rref{fundp}, as well as \rref{cd2}, is
\be
\hat c ={\hat P}_1 T, \qquad \hat d ={\hat P}_2 T.
\label{cdsol}
\ee
In this case their action is given by equation \rref{pt}, and from
\rref{ppt} they
are idempotent ${\hat c}^2={\hat d}^2={\rm id}$.

All four terms in the second of \rref{fundp} are then proportional
\be
{\hat P}_1 \hat d =e^{i\hbar}\hat d {{\hat P}_1}^{-1}=\hat c
{{\hat P}_2}^{-1}= e^{i\hbar}{\hat P}_2 \hat c
\label{sol2}
\ee
and act as
\be
{\hat P}_1 \hat d \psi(b)=
 e^{i\hbar (1+b)}\psi(-b-1)
\label{actsol2}
\ee
Powers of the operator matrices \rref{matr1} in this case are then expressed by
\be
{\hat U_1}^n=
\left(\begin{array}{clcr}{{\hat P}_1}^n&n{{\hat P}_1}^n T\\0
&{{\hat P}_1}^{-n}\end{array}\right), \qquad
\label{matr1n}
{\hat U_2}^m=
\left(\begin{array}{clcr}{{\hat P}_2}^m&m{{\hat P}_2}^m T\\0&{
{\hat P}_2}^{-m}\end{array}\right)
\ee
and satisfy
\be
{\hat U_1}^n {\hat U_2}^m = e^{inm \hbar}{\hat U_2}^m {\hat U_1}^n
\ee
for $m, n ~\epsilon ~\IZ$, where
\be
{\hat U_1}^n{\hat U_2}^m= \left(\begin{array}{clcr}
{{\hat P}_1}^n{{\hat P}_2}^m&(m+n){{\hat P}_1}^n{{\hat P}_2}^m T
\\0&{{\hat P}_1}^{-n}{{\hat P}_2}^{-m}
\end{array}\right).
\label{produu}
\ee
The internal relations between the elements of ${{\hat U}_1}^n$
or ${{\hat U}_2}^m$ or the product ${{\hat U}_1}^n {{\hat U}_2}^m$
are the same as those for ${\hat U}_1, {\hat U}_2$ \rref{cd2},
as can be shown by repeated use of the first of \rref{fundp} and
\rref{ppt}.

\section{Quasi--Modular transformations \label{sec4}}

It is known that in (2+1)--dimensional gravity with
topology $\IR \times T^2$,
the set of equivalence classes of ``large'' diffeomorphisms (modulo
diffeomorphisms that can be deformed to the identity) are generated
by two independent Dehn twists which act on the classical holonomies
$U_1, U_2$ as
\begin{eqnarray}
&S&: U_1\rightarrow U_2, \qquad
     U_2\rightarrow{U_1}^{-1}\nonumber\\
&T&: U_1\rightarrow U_1 U_2,\qquad
     U_2\rightarrow U_2
\label{modu}
\end{eqnarray}
which leave invariant the commutator $U_1 U_2 { U_1}^{-1} {U_2}^{-1}$
and satisfy $(ST)^3={\rm id}, S^4={\rm id}$.
For a discussion of the modular group in (2+1)--dimensional gravity
see \cite{cn2} and \cite{carlbk}.

In this section we attempt to implement the transformations \rref{modu} on
the solutions found in Sections \ref{sec3} and \ref{sec4}.
One approach to implementing the modular transformations in the quantum
theory is by acting on parameters (for a slightly different approach
see \cite{jrmath}). For the algebraic solution in Section
\ref{sec2} it can
be checked that the transformations
\begin{eqnarray}
S&:& \a_1\rightarrow \a_2, \qquad
     \a_2\rightarrow{\a_1}^{-1}\nonumber\\
 &:& \b_1\rightarrow \b_2, \qquad \b_2\rightarrow
{\a_1}^{-2}\b_1\nonumber\\
T&:& \a_1\rightarrow q^{-\um}\a_1\a_2, \qquad
     \a_2\rightarrow\a_2\nonumber\\
 &:& \b_1\rightarrow q^{-\um}\b_1\a_2^{-1}=q^{-\um}\a_1\b_2, \qquad
\b_2\rightarrow \b_2
\label{moduuu}
\end{eqnarray}
do NOT give exactly \rref{modu} but do preserve the
fundamental relation \rref{fundv} and
satisfy $(ST)^3={\rm id}, S^4={\rm id}$. Moreover, they generate new operator matrices
{\it of the same type}, i.e. with the same internal relations.

The corresponding transformations for the operator representations of Section
\ref{sec3} are
\begin{eqnarray}
S&:&{\hat P}_1\rightarrow {\hat P}_2,\quad
    {\hat P}_2\rightarrow {{\hat P}_1}^{-1},\nonumber\\
T&:&{\hat P}_1\rightarrow e^{-{i \hbar} \over 2}{\hat P}_1{\hat P}_2,\quad
{{\hat P}_1}^{-1}\rightarrow (e^{-{i \hbar} \over 2}{\hat P}_1{\hat P}_2)^{-1}=
e^{-{i \hbar} \over 2}{{\hat P}_1}^{-1}{{\hat P}_2}^{-1},\quad
 {\hat P}_2\rightarrow {\hat P}_2,
\label{modp}
\end{eqnarray}
which leave \rref{fundp} invariant.

The matrices \rref{matr1}, with the solution \rref{cdsol}, that we are considering
transform, from \rref{modp}  as
\begin{eqnarray}
&S&:\hat U_1\rightarrow \tilde {\hat U_1}=\hat U_2, \qquad
    \hat U_2\rightarrow \tilde {\hat U_2}= \left(\begin{array}{clcr}
    {{\hat P}_1}^{-1}& T {\hat P}_1\\0&{\hat P}_1 \end{array}\right) \nonumber\\
&T&:\hat U_1\rightarrow\tilde {\hat U_1}=e^{-{i \hbar \over 2}}\left(
\begin{array}{clcr} {\hat P}_1{\hat P}_2&{\hat P}_1{\hat P}_2 T\\
0&{{\hat P}_1}^{-1}{{\hat P}_2}^{-1}
\end{array}\right),\qquad
    \hat U_2\rightarrow \tilde{\hat U_2}=\hat U_2
\label{moduu}
\end{eqnarray}
with $(ST)^3={\rm id}, S^4={\rm id}$, and give rise to operator
matrices
$\tilde {\hat U_1},\tilde {\hat U_2}$ {\it of
the same type}, i.e. having internal relations like \rref{cd2}
and, moreover,
satisfying the fundamental relation \rref{fund1}.

\section{Discussion \label{sec5}}

We have found both algebraic solutions and operator representations
of diagonal and upper--triangular operator matrices ${\hat U}_1,
{\hat U}_2$ that satisfy the fundamental relation \rref{fund}. They
furnish a generalisation of previous approaches to the quantisation
of (2+1)--dimensional gravity which only use the traces
\cite{NRZ,NR1}, but are still consistent with those approaches. The
difference between the diagonal and upper-triangular cases only
shows up very indirectly when the traces are used. The matrix
variables give a much clearer insight into the structure of the
phase space, both before and after quantisation. It is interesting
to note that in the quantum theory the diagonal and upper
triangular sectors play an equally important role, unlike the
classical case where the latter sector is of lower dimension
\cite{jrmod}.

In principle the upper--triangular algebraic solutions (and
their operator representations in Section \ref{sec3}) could be
used
for an alternative description of (2+1)--dimensional gravity.
Although we introduce extra variables there are extra relations
between them and the counting of degrees of freedom is the same.
As described in Section \ref{sec1} each $(\pm)$ sector has one
degree of freedom (the variables $\hat r_1, \hat r_2$ satisfying
the commutator \rref{dc1}). For the algebraic solution of
Section \ref{sec2}, there are six variables $\a_i,\b_i,\g_i, i=1,2$
which satisfy \rref{ab1} (2 relations), \rref{pair1} (2 relations),
and \rref{hom1} (1 independent relation). The other $(+{\rm or} -)$
sector would have similar relations with $q$ replaced by $q^{-1}$,
and its variables would commute with those of the first sector.

The diagonal solutions are themselves interesting and transform as
\rref{modu} apart from phases under the modular group action. The
entries of each diagonal matrix commute internally but the
entries from different matrices obey standard $q$--group
relations.

The upper--triangular case is unusual and interesting in that the
matrices cannot be diagonalised because they have non--commuting
entries.  For consistency with \rref{fund}, two types of
non--commutativity are required. The first -- non--commutativity
between elements from different holonomies -- follows from the
fundamental relation -- a deformation of the "two commuting
holonomies" statement. This is just the non--commutativity
encountered when we quantise a classical system, with some $q$ or
$\hbar$ parameter, although not all the relations are standard
$q$--group relations. The second -- non--commutativity between {\it
internal} operator entries of each matrix -- may be an example of
the non--commutativity often encountered in fermionic systems. In
fact this feature persists in the classical limit $q \to 1$. These
non--trivial internal relations \rref{int} are also not standard
$q$--group commutation relations. For example, in Section
\ref{sec3}, if we compare the action of $c{\hat P}_1$ and ${\hat
P}_1 c$, from \rref{pt} and \rref{actp12}, we find that they can
only be proportional (equal up to a phase) if ${\hat P}_1= {\rm id}.$

This combined structure has interesting mathematical features which are
further explored in \cite{jrmath}. It is perhaps most closely related to
the notion of quantum braided groups \cite{maj}.

The {\it quasi--modular transformations} \rref{moduu} generate
matrices of the same type i.e. satisfying the fundamental
relation \rref{fund1} and with internal relations of the form \rref{cd2}.

\section*{Acknowledgments}
This work was supported in part by the European Commission HCM programme
CHRX-CT93-0362, the European Commission TMR programme
ERBFMRX-CT96-0045, INFN Iniziativa Specifica FI41, and the
Programa de Financiamento Plurianual of the

\noindent Funda\c{c}\~{a}o para a Ci\^{e}ncia e a Tecnologia (FCT).

\end{document}